\documentclass[twocolumn,superscriptaddress,floats,prd,nofootinbib,showpacs]{revtex4-1}

\usepackage{amssymb,amsmath,amssymb,amsfonts,amsthm,stmaryrd,mathrsfs,physics,bbm,mathtools,lipsum}
\usepackage{graphicx}
\usepackage[T1]{fontenc}
\usepackage{enumerate} % advanced enumerate environment
\usepackage{color} % for colored text
\usepackage{graphpap} % for numbered coordinate grid
\usepackage{xcolor}
\usepackage{bm} 
\usepackage{hyperref} % for HyperTeX cross-referencing
\usepackage[normalem]{ulem}

\newcommand{\sgn}{\mathrm{sgn}}

\begin{document}

\title{Black hole image encoding quantum gravity information}

\author{Cong Zhang}
\email{czhang(AT)fuw.edu.pl}
\affiliation{Department Physik, Institut f\"ur Quantengravitation, Theoretische Physik III, Friedrich-Alexander-Universit\"at Erlangen-N\"urnberg, Staudtstra{\ss}e 7/B2, 91058 Erlangen, Germany}

\author{Yongge  Ma}
\email{mayg(AT)bnu.edu.cn}
\affiliation{Department of Physics, Beijing Normal University, Beijing 100875, China}

\author{Jinsong Yang}
\email{jsyang(AT)gzu.edu.cn}
\affiliation{School of Physics, Guizhou University, Guiyang 550025, China}

\begin{abstract}
The quantum extension of the Kruskal spacetime indicates the existence of a companion black hole in the universe earlier than ours. It is shown that the radiations from the companion black hole can enter its horizon, pass through the deep Planck region, and show up from the white hole in our universe. These radiations inlay extra bright rings in the image of the black hole in our universe, and some of these rings appear distinctly in the shadow region. Therefore, the image of the black hole observed by us encodes the information of quantum gravity. The positions and widths of the bright rings are predicted precisely. The predictive values for supermassive black holes are universal for a quite general class of quantum modified spacetime with the scenario of black hole to white hole transition. Thus, our result opens a new experimental window to test this prediction of quantum gravity. 
\end{abstract}

%\keywords{Black hole, loop quantum gravity, singularity resolution }
\maketitle
%\tableofcontents

\section{introduction}

How to unify  quantum mechanics and general relativity remains one of the most challenging topics in modern physics. Although various theoretical approaches have been proposed towards this goal (e.g., \cite{polchinski1998string,ashtekar2015general,thiemann2008modern}), the lack of experimental data poses a significant obstacle to constructing a viable theory of quantum gravity (QG). Hence, to explore novel experimental phenomena of QG is of utmost importance. 
%It was argued by Dyson that that detecting a graviton, the hypothetical elementary particle that mediates the force of gravity, might be unrealistic \cite{rothman2006can,dyson2013graviton}. 
It has been studied to detect QG effects through, for instance, entanglement between massive particles \cite{marletto2017gravitationally}, Lorentzian invariance violation \cite{collins2004lorentz}, and other QG phenomenology \cite{rothman2006can,dyson2013graviton,addazi2022quantum}.

Since the gravitational waves were detected by LIGO/Virgo \cite{abbott2016observation} and the black hole (BH) images were photographed by the Event Horizon Telescope (EHT)  \cite{collaboration2019first,akiyama2022first}, how to detect QG effects by applying these observations has attracted the extensive attention \cite{liu2020shadow,PhysRevLett.126.181301,Afrin:2022ztr,Vagnozzi:2022moj,yang2022loop}. However, so far those studies were limited to the classical region outside the BH horizon, and thus the QG effects would be  minuscule. To obtain significant information about QG, one would expect to observe the signals from highly quantum regions inside the BH horizon.
This is impossible in classical physics due to the existence of the horizon and singularity. However, the quantum extension of a BH spacetime indicates that the classical singularity can be resolved and there are a series of universes other than ours where companion BHs exist  \cite{ashtekar2018quantum,husain2022quantum,lewandowski2022quantum}. This provides an opportunity to observe the light signals that enter the horizon of a companion BH in the universe earlier than ours, travel through the highly quantum region, and occur in the image of the BH in our universe. In this paper we willl demonstrates that these lights indeed produce extra bright rings in the BH image, and some of them appear in the classical shadow region. The positions and widths of  these bright rings can be precisely predicted, and they are universal for a broad class of quantum modified spacetimes, where a transition of BH to white hole (WH) occurs. Our results open a new experimental window to test this prediction of QG. Moreover, the next-to-leading order of the predictive values is model-dependent and thus encodes the information to distinguish different candidate theories.

\begin{figure}
\centering
\includegraphics[width=0.4\textwidth]{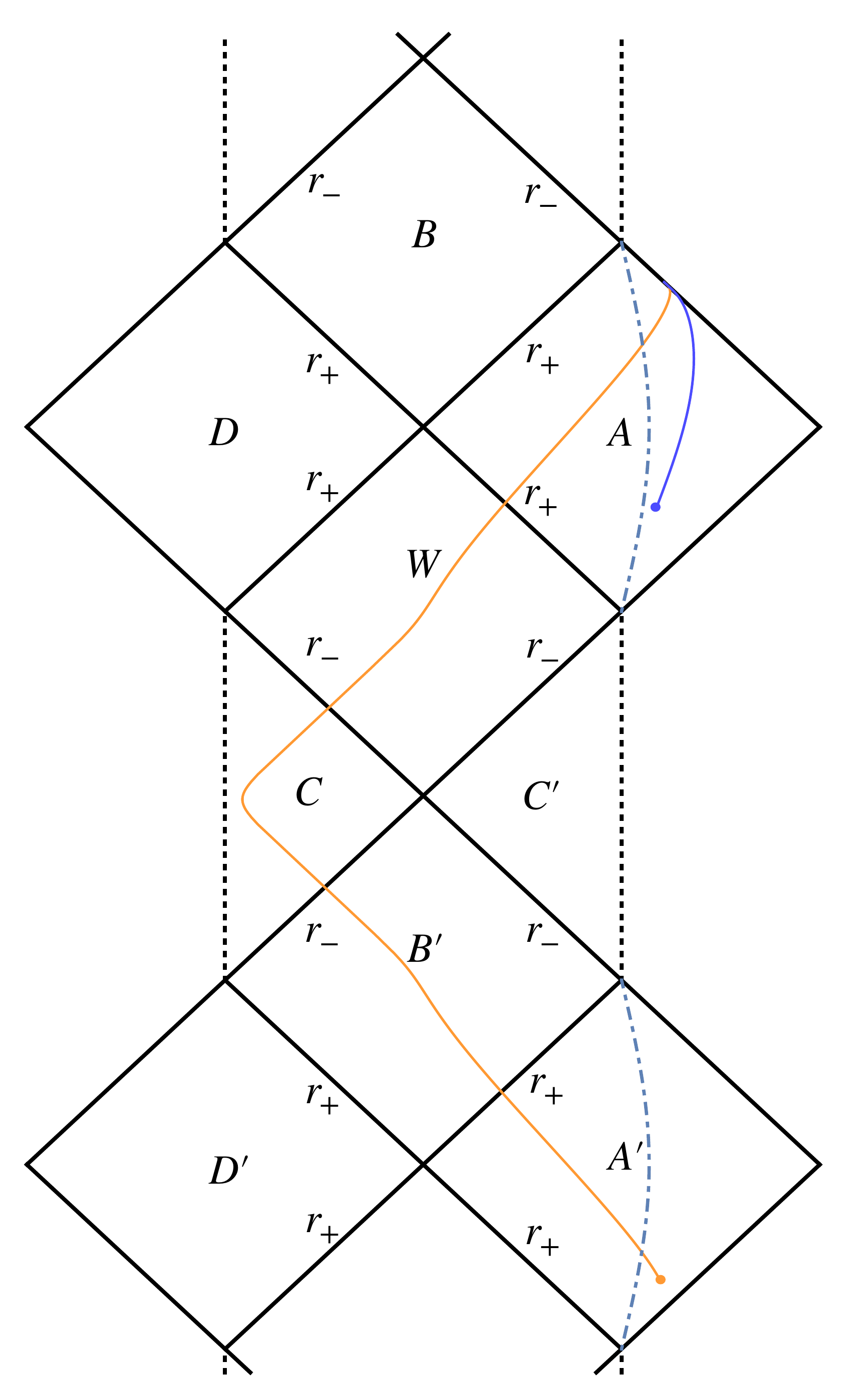}
\caption{The Penrose diagram of the quantum extended BH spacetime: The golden and blue lines are two null geodesics emitted from the thin disks in $A'$ and $A$ respectively and observed by the observer in Region $A$ simultaneously. The dash-dotted blue line illustrates the location of the photon sphere.}\label{fig:penrose}
\end{figure}

A few quantum extensions of the Kruskal spacetime have been proposed in the study of loop quantum gravity (LQG) \cite{modesto2004disappearance,gambini2008black,agullo2008black,hossenfelder2010model,haggard2015quantum,corichi2016loop,ashtekar2018quantum,zhang2020loop,gan2020towards,gambini2020spherically,giesel2021nonsingular,han2022improved,husain2022quantum,lewandowski2022quantum}. 
While the model in \cite{lewandowski2022quantum} will be employed in the following  calculation, our analysis is valid for all these quantum spacetimes where a BH to WH transition occurs. The metric of the LQG modified spherically symmetric  spacetime reads
 \begin{equation}\label{eq:metric}
\dd s^2=-f(r)\dd t^2+f(r)^{-1}\dd r^2+r^2(\dd\theta^2+\sin^2\theta\dd\varphi^2),
\end{equation}
with
\begin{equation}
\begin{aligned}
f(r)=1-\frac{2M}{r}+\frac{\alpha M^2}{r^4},\quad  \alpha=16\sqrt{3}\pi\gamma^3\ell_p^2
\end{aligned}
\end{equation}
where the  geometric unit with $G=1=c$ is chosen such  that $\ell_p=\sqrt{\hbar}$ is the Planck length and $\gamma$ denotes the Barbero-Immirzi parameter \cite{BarberoG:1994eia}.  This modified spacetime was derived from two independent approaches in LQG \cite{Husain:2022gwp,lewandowski2022quantum}  and attracted considerable attention \cite{Giesel:2022rxi,Han:2022rsx,Han:2023wxg,Bobula:2023kbo}.   
%Clearly, the modification term is purely quantum since as $\hbar\to 0$ the metric returns to the Schwarzschild one. 
For realistic consideration, we assume that the BH is massive, i.e., $M\gg 4\sqrt{\alpha}/(3\sqrt{3})$. In this case, $f(r)$ has two real roots $r_\pm>0$, corresponding to the two horizons in the maximally  extended spacetime \cite{munch2021causal,lewandowski2022quantum}, whose Penrose diagram is shown in  Fig. \ref{fig:penrose}. 

Consider a static observer in the asymptotically flat region $A$ near the future null infinity to detect the image of the companion BH $B'$ and the BH $B$ illuminated by  distant, uniform, isotropically emitting screens surrounding them. Two null geodesics emitted by the screens surrounding $B$ and $B'$ are plotted by blue and golden lines respectively in Fig.  \ref{fig:penrose}.  The image photographed by the observer should be the overlay of the images resulting from the ``blue'' and ``golden'' null geodesics. As the image produced by the geodesics of  the blue lines has been given in \cite{yang2022loop}, in the current work, we focus on the image from the golden geodesics and study their overlay. To this end, we first analyze the behavior of null geodesics in the spacetime with metric \eqref{eq:metric}. Then, we consider the model of thin disk emission to study the contributions of the golden geodesics to the BH image. 
%Our key prediction is a consequence of \eqref{eq:phiouttot} where the leading-order terms are valid for all quantum modified spactimes taking the Schwarzschild as the classical limit, as shown by our discussion.   

\section{null geodesics in the quantum modified spacetime}
With the Eddington–Finkelstein advanced coordinate $(v_+,r,\theta,\varphi)$ in $A'\cup B'\cup C$ and the retarded coordinate $(v_-,r,\theta,\varphi)$ in $C\cup W\cup A$, %\cite{hawking2023large,lewandowski2022quantum}, 
the conservation equations of energy and angular momentum together with the vanishing norm of the null geodesics imply
\begin{eqnarray}
\left(\frac{\dd u}{\dd \phi}\right)^2&=&-\alpha M^2 u^6+2 M u^3-u^2+\frac{1}{b^2}\equiv G(u,b),\label{eq:nullgeo1} \\ 
\frac{\dd v_\pm}{\dd u}&=&\frac{\pm b}{\sqrt{G(u,b)}(1+b\sqrt{G(u,b)})},\label{eq:nullgeo2}
\end{eqnarray}
where $u\equiv 1/r$, $\phi\in (0,\infty)$ is the azimuthal angle in the orbit plane, and $b$, the impact parameter given by the ratio of the angular momentum and energy, is a constant of motion and chosen to be positive. 
Note that the affine parameter has been canceled in Eqs. \eqref{eq:nullgeo1} and \eqref{eq:nullgeo2} for the geodesic trajectories.  

Equation \eqref{eq:nullgeo1} ensures  $G(u,b)\geq 0$ along the geodesics and implies that the turning points occur at the roots $u$ of  $G(u,b)$. It is easy to check that  $G(u,b)=0$ allows at most four real roots and two of them always exist. One is negative and thus unphysically,  while  the other, denoted by $u_0$, is positive and satisfies  $u_0>1/r_-$. Let $b_c$ denote the value of the impact parameter for which the remaining two roots are a double root $u_{\rm ph}$. 
Then, $u_{\rm ph}$ and $b_c$ can  be obtained by solving equations $\partial_uG(u_{\rm ph},b_c)=0$ and $ G(u_{\rm ph},b_c)=0$. 
It turns out that $r_{\rm ph}=1/u_{\rm ph}>r_+$ characterizes the photon sphere and $r(\phi)=r_{\rm ph}$ represents unstable null geodesics. They are plotted as the dash-dotted blue lines in Fig. \ref{fig:penrose}. The cases with $b>b_c$ are not considered in the current work since  these null geodesics either lie entirely outside the horizon $r_+$ or are confined in some finite radius.  For the case of $b<b_c$ where the remaining two roots are both complex, the null geodesics start from $r=\infty$, cross the two horizons and turn back at $1/u_0$, as the golden line in Fig. \ref{fig:penrose}.   This fact can be confirmed by checking that $v_\pm(u)$ is always finite along the geodesic by integrating  Eq. \eqref{eq:nullgeo2}.

\section{BH images with thin disk emission}
\begin{figure}
\includegraphics[width=0.45\textwidth]{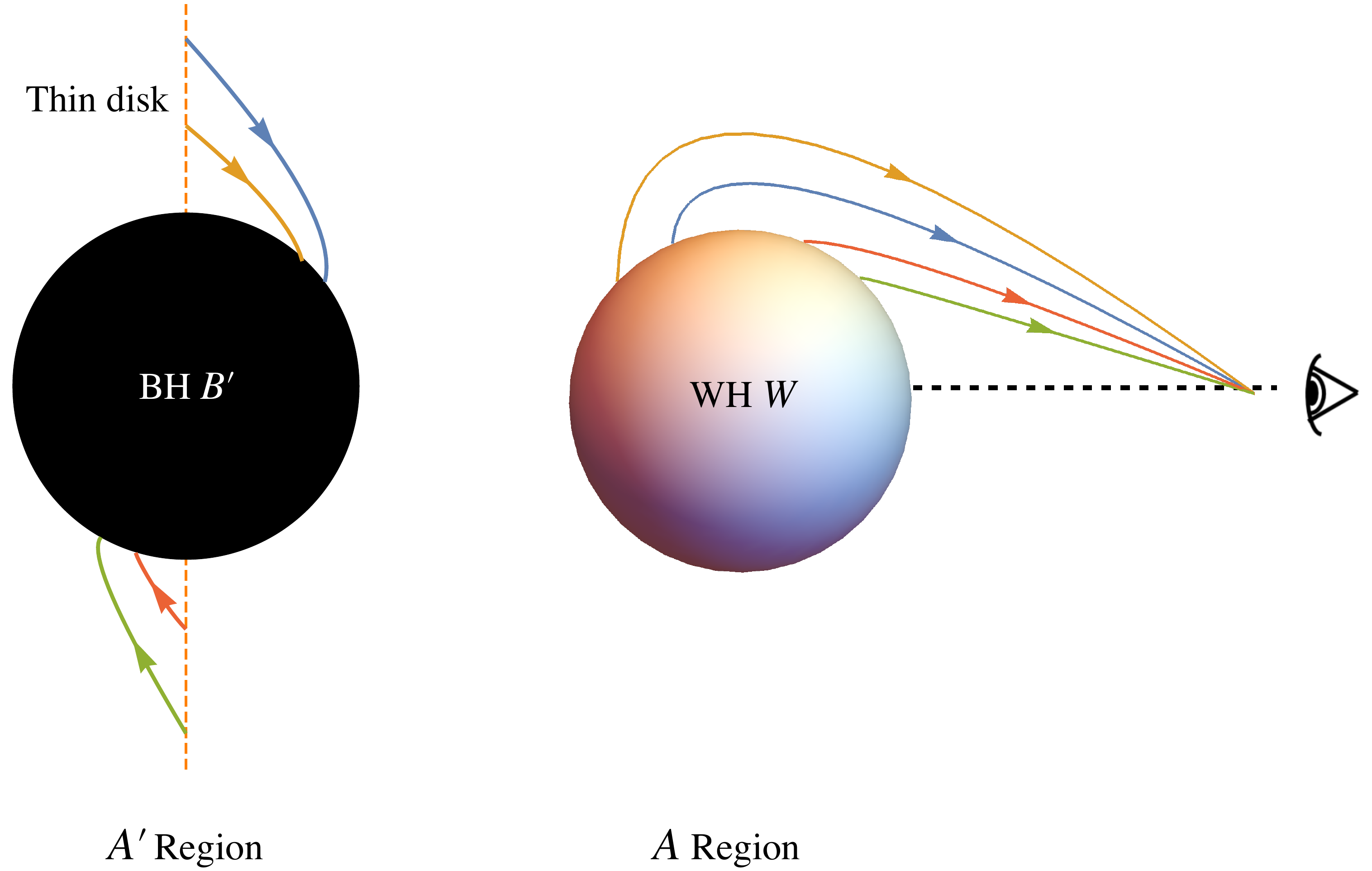}
\caption{The schematic of our model where the dashed orange line shows the thin disk and the curves with arrows depict the null geodesics with various impact parameters. 
}\label{fig:nullgeos}
\end{figure}

To investigate the contribution of the geodesics with $b<b_c$ to the BH image, we consider a model where the emissions originate from an optically and geometrically thin disk which stays at rest outside the BH $B'$ and emits isotropically in the rest frame of the static observers \cite{gralla2019black,peng2021observational}. 
The disk lies in the equatorial plane, and the observer in region $A$ faces the north pole of the WH $W$, as shown in Fig. \ref{fig:nullgeos}.
%To understand the image of the BH $B'$ illuminated by the disk, we trace the null geodesics from the eye of the observer  backwards towards the region near $B'$.
In this model, the relevant parts of each geodesic are those in Regions $A$ and $A'$. The trajectories $u(\phi)$ of the geodesics can be solved from Eq. \eqref{eq:nullgeo1} numerically (see \cite{github} for our numerical codes). Here the initial condition is $\phi=0$ for $u=0$ in Region $A$, since the lights captured by the observer facing the north pole  are parallel.

Let $I_\nu^{\rm em}$ denote the specific intensity of light from the thin disk. It is assumed to depend only on the radial $r$, i.e., $I_{\nu}^{\rm em}(r)=I(r)$  for all $\nu$, where $\nu$ is the frequency in a static frame. Consider the radiation emitted from a radius $r_o$ in $A'$. 
Due to the symmetry of the spacetime, the history of the radiation after the turning point is the time reverse of its history before the turning point (see  Fig. \ref{fig:penrose}). Therefore, when it arrives at the same radius $r_o$ in $A$, the light ray carries the same  specific intensity and frequency as it has at $r_o$ in $A'$. Since the remaining journey of the light entirely lies in $A$, we can employ the results in \cite{gralla2019black,peng2021influence} to conclude that the radiation will be received at some frequency $\nu'$ with the specific intensity 
$I_{\nu'}^{\rm obs}=f(r_o)^{3/2}I(r_o).$ Consequently, the integrand intensity $I=\int I\dd\nu$ scales as $f^2$, i.e., 
$I^{\rm obs}=f(r_o)^2 I(r_o).$ To understand the image of $B'$ illuminated by the disk, we trace the null geodesics from the observer backwards towards the region near $B'$ as in \cite{gralla2019black}.  
Since the thin disk is assumed to be optically and geometrically thin, the light ray may intersect with the thin disk many times and pick up the brightness repeatedly.  Hence, the observed light intensity is a sum of the intensities from each intersection
i.e., $$I^{\rm obs}=\sum_{m}f^2I\big|_{r=r_m(b)},$$ where $r_m(b)$ is the radial coordinate of the $m$th intersection of the light rays with the disk plane outside the horizon.

\begin{figure}
\centering
\includegraphics[width=0.5\textwidth]{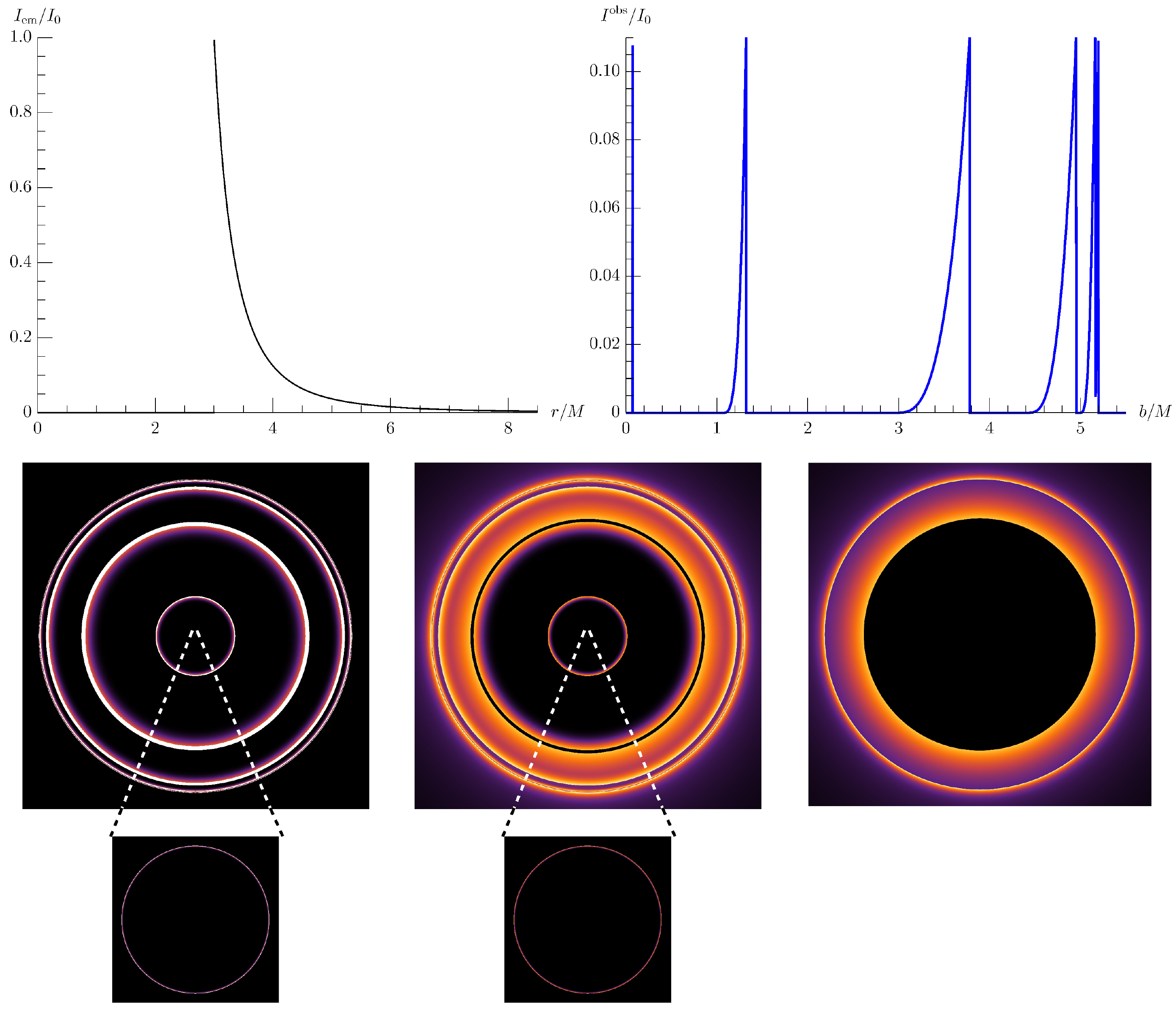}
\caption{The  BH images: The emitting intensity  of the thin disk surrounding the BH $B'$ peaks at the photon sphere $r_{\rm ph}\approx 3M$ with $M=243.747\sqrt{\alpha}$ (the top-left panel). The observed intensity is shown in the top-right panel, and its image is given in the bottom-left panel. The image of $B'$ comprises five distinguishable bright rings. The corresponding values of $(b_n/M,b_n'/M)$ for $n=1,2,3,4$ and $5$ are $(0.072,0.076),(1.077,1.512),(2.974,4.373), (4.414,5.146)$ and $(4.995,5.194)$, respectively. The bottom-right panel gives the image of the BH $B$, also surrounded by a thin disk in the equatorial plane with the same emission as that  of $B'$. The overlay of  images of $B$ and $B'$ is shown in the bottom-middle panel.    
}\label{fig:shadow}
\end{figure}

An example of the  images of the companion BH $B'$ and the BH $B$ is shown in Fig. \ref{fig:shadow}. It illustrates that the image of $B'$ comprises five bright rings. The first three appear distinctly in the shadow region of the image of B, while the last two are inlaid as bright rings in the illuminated region. Note that the gap between the screen and the horizon results in the differences between $b'_n/M$ and the outer edges of the nth bright ring. These bright rings carry the information of QG, since they  come from the light rays emitting from $B'$ and traveling  through the deep Planck region $C$. 

\section{positions and widths of the bright rings}
\begin{figure}
\centering
\includegraphics[width=0.45\textwidth]{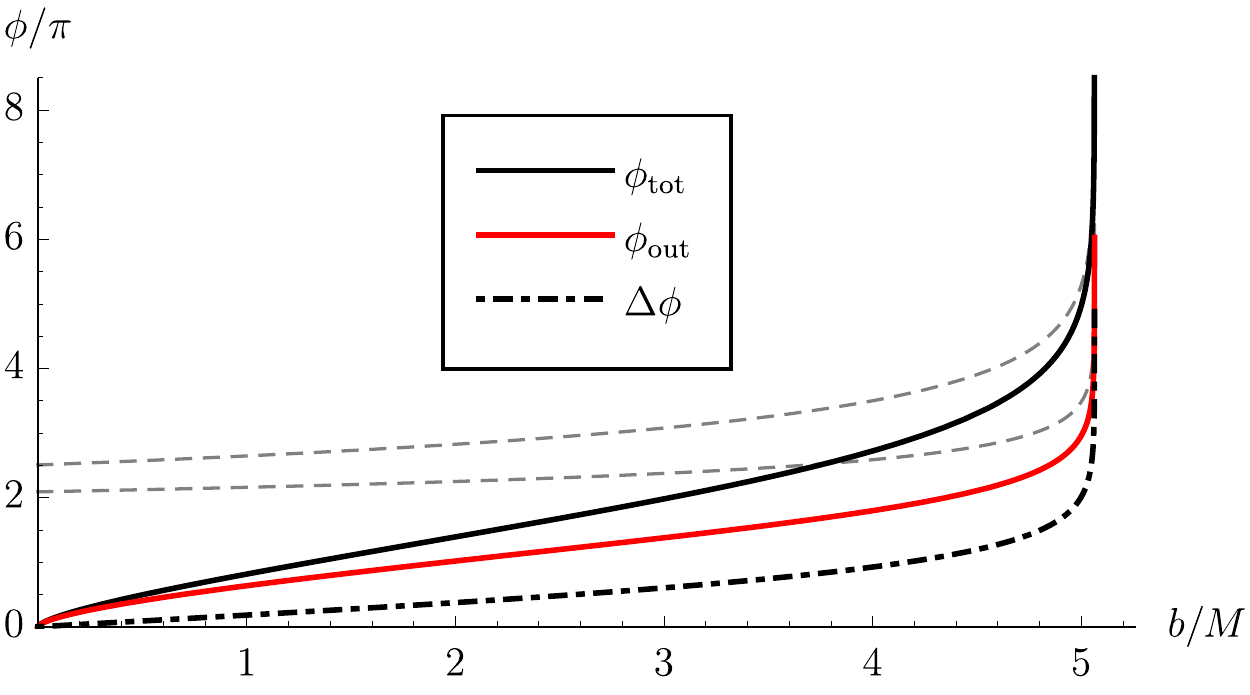}
\caption{The dependence of $\phi_{\rm out}$, $\phi_{\rm tot}$ and $\Delta\phi=\phi_{\rm tot}-\phi_{\rm out}$ on $b/M$: The dashed gray lines are obtained by the analytical  approximation (see Appendix \ref{app:B}). The parameters are chosen as $M=243.75\sqrt{\alpha}$, $\hbar=1$ and $\gamma=0.2375$. 
}\label{fig:phivalues}
\end{figure}

The bright rings are the consequence of the intersections of the null geodesics with the thin disk.
Whether a light ray intersects the disk is determined by the two azimuthal angles in orbital plane as the ray arrives at the horizon $r_+$ and infinity in $A'$, denoted by $\phi_{\rm out}(b)$ and $\phi_{\rm tot}(b)$ respectively. Their values can be calculated numerically.  According to the result shown in Fig. \ref{fig:phivalues}, as $b$ increases from $0$ to $b_c$, $\phi_{\rm tot}(b)$, $\phi_{\rm out}(b)$, and the difference  $ \Delta\phi(b)\equiv \phi_{\rm tot}(b)-\phi_{\rm out}(b)$ increase.  To see the implication of this result to the BH image, let us start with $b=0$ and increase its value. As shown in Fig.  \ref{fig:phivalues}, both $\phi_{\rm out}(b)$ and $\phi_{\rm tot}(b)$ are initially smaller than $\pi/2$,and hence these rays can not intersect the thin disk. 
Subsequently, as $b$ increases to some value $b_1$, $\phi_{\rm tot}(b)$ reaches $\pi/2$ at first, so this ray and those right after could get the brightness from the thin disk. 
As $b$ increases to some value $b_1'$, $\phi_{\rm out}(b_1')$ arrives at $\pi/2$ while $\phi_{\rm tot}(b_1')$ is smaller than $3\pi/2$, so that 
this ray and those right after cannot intersect with the disk. Hence those rays with impact parameters $b\in (b_1,b_1')$ can intersect with the thin disk. It should be noted that in practical case the thin disk has a finite size and there is a distance between the disk and the horizon, as in the example shown in Fig. \ref{fig:shadow}. Then not all of the light rays with $b\in (b_1,b_1')$ could be lightened by the disk. 
Those lightened rays can be finally observed in Region $A$ as a light ring in the BH image.  The analysis similar to the above can continue as the impact parameter increases. As shown in Fig. \ref{fig:phivalues}, there exists certain value $b_\pi$ whose corresponding light ray satisfies $\Delta\phi(b_\pi)=\pi$. This is the critical point such that all of the rays with $b>b_\pi$ will intersect the equatorial plane anyway. Thus, the bright image of these rays could become continuous, in spite of that these rays may contribute bright rings in the practical cases as in Fig. \ref{fig:shadow}.

The above discussions indicates that we can introduce two series of the impact parameters $b_n$ and $b_n'$ satisfying $\phi_{\rm tot}(b_n)=(2n-1)\pi/2$  and $\phi_{\rm out}(b_n')=(2n-1)\pi/2$ respectively to specify the edges of the bright rings. For $M\gg \sqrt{\alpha}$, the value of $b_n$ and $b_n'$ can be analytically obtained by solving Eq. \eqref{eq:nullgeo1} with $b=\beta M$. To this end, we may apply the method of Matched Asymptotic Expansions \cite{lagerstrom2013matched,PhysRevD.95.064009,Hou_2022}, and obtain (see Appendix \ref{app:A} for the detailed derivations)
 \begin{equation}\label{eq:phiouttot}
\begin{aligned}
\frac{\phi_{\rm tot}(\beta M)}{2}&=\int_0^\infty \frac{\dd y}{\sqrt{2y^3-y^2+\beta^{-2}}}\\
&-\frac{\sqrt[6]{4\pi^3 }\Gamma \left(\frac{5}{6}\right)}{\Gamma \left(\frac{1}{3}\right)}\frac{\sqrt[6]{\alpha}}{\sqrt[3]{M}}+O(\frac{\alpha^{1/3}}{M^{2/3}}),\\
\Delta\phi(\beta M)&=\int_0^{\frac{1}{2}} \frac{\dd y}{\sqrt{2y^3-y^2+\beta^{-2}}}+O(\frac{\sqrt{\alpha}}{M}),
\end{aligned}
\end{equation}
Note that the leading order contributions to $\phi_{\rm tot}(\beta M)/2$ and $\Delta\phi(\beta M)$ coincide with the changes of the azimuthal angles for a light traveling from $r=\infty$ to the singularity and to the horizon in the Schwarzschild spacetime, respectively \cite{Luminet:1979nyg}, since the quantum spacetime takes the Schwarzschild one as its classical limit. This implies that the leading order terms of Eq. \eqref{eq:phiouttot} may  be valid for a general class of quantum modified metrics with this classical limit. It is indeed the case. As shown by the detail calculation in  Appendix \ref{app:A}, the following conditions on the quantum-modified metric \eqref{eq:metric} with a general $f(r)$ can insure that the leading order terms of Eq. \eqref{eq:phiouttot} remain true :
\begin{itemize}
\item[(i)]  In the classical region with $r>M$, the quantum modified metric returns to the Schwarzschild one, i.e., with the scaling $r=wM$ and $p>0$, we have
\begin{equation}
f(w M)=1-\frac{2}{w}+O((\sqrt{\alpha}/M)^{p}).\label{eq:classicallim1}
\end{equation}
\item[(ii)] In the quantum region $r\sim (\alpha M)^{1/3}$ of the Schwarzschild spacetime, where the Kretschmann scalar approaches the Planck value, the quantum corrections to the metric are of the order $\left(M^2/\alpha\right)^{1/3}$, which manifests as the term $2M/r$, i.e., with scaling $r=w(\alpha M)^{1/3}$ and introducing a function $F(w)$, we have:
\begin{equation}
f(w(\alpha M)^{1/3})=\left(\frac{M^2}{\alpha}\right)^{1/3}F(w)+O((\sqrt{\alpha}/M)^{0}),\label{eq:classicallim2}
\end{equation}
\item[(iii)] The turning point of the null geodesic with  $b=\beta M$, caused by the quantum correction, is located in the quantum region $r\sim (\alpha M)^{1/3}$.
\end{itemize}
The function $F(w)$ introduced in condition (ii) encodes the quantum correction to the metric, and its specific form will depend on the particular quantum black hole model under consideration. By examining the higher-order terms in \eqref{eq:phiouttot}, which involve $F(w)$, one can distinguish different models of quantum black holes. This explores how the quantum modified metric affects the BH images.

\begin{figure}
\centering
\includegraphics[width=0.4\textwidth]{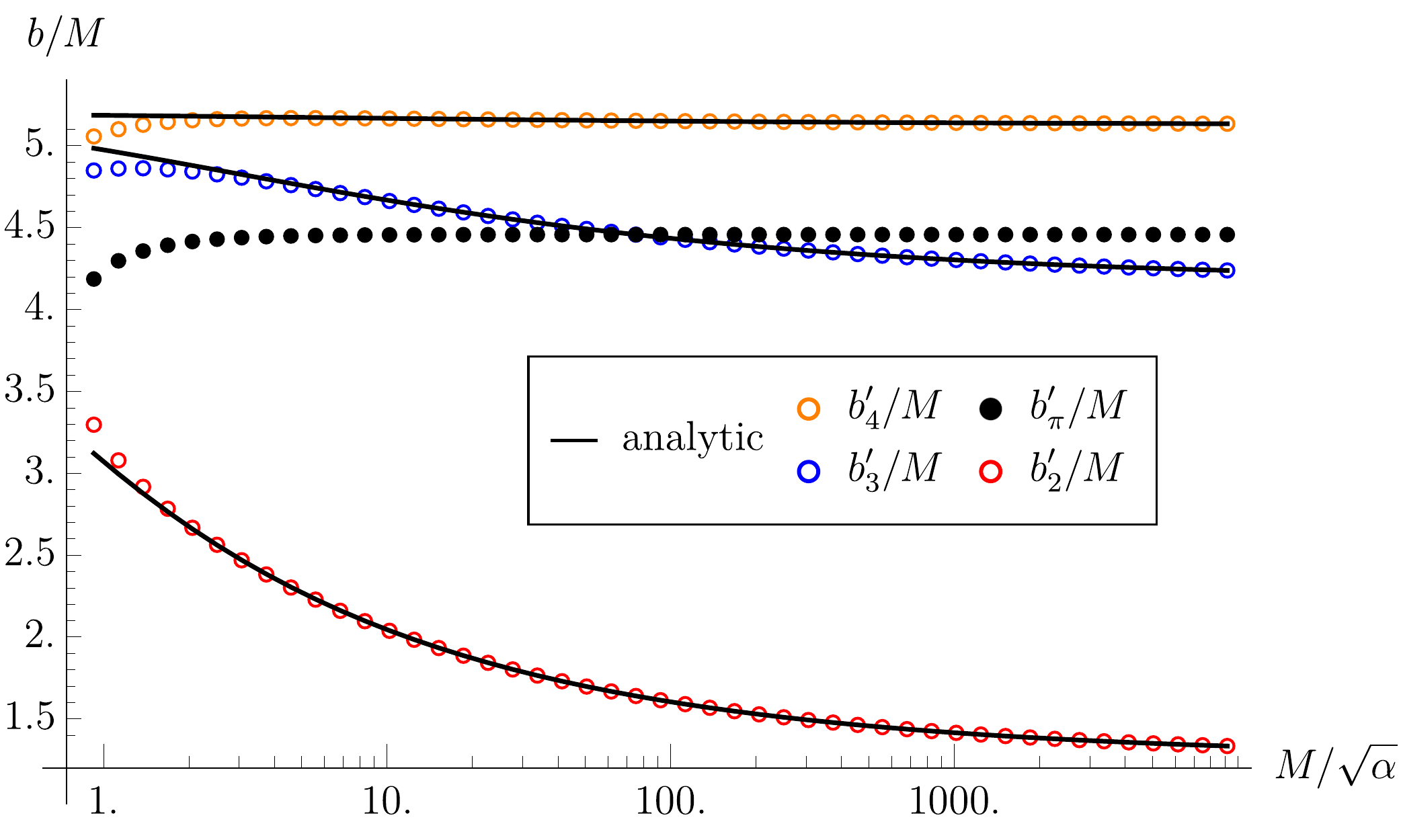}
\caption{The comparison of the analytic and numerical values of $b'_n/M$ for $n=2,3,4$ and $b_\pi/M$: The analytic approximations obtained by applying Eq. \eqref{eq:phiouttot} are plotted by black lines.
}\label{fig:cribs}
\end{figure}

%By neglecting the higher order terms of $\sqrt\alpha/M$, the values of $b_n/M$ and $b_n'/M$ as \textcolor{red}{$M \to \infty$} can be solved by applying Eq. \eqref{eq:phiouttot}. Thus, 
The intervals $(b_n/M, b_n'/M)$ to specify the bright rings as $M \to \infty$  can be obtained by applying Eq. \eqref{eq:phiouttot} and ignoring higher order terms of $\sqrt\alpha/M$. The limits of $(b_n/M, b_n'/M)$ are $(0.0427,0.0445)$, $(0.9242,1.2620)$, $(2.7927,4.1766)$, $(4.3175,5.1285)$ and $(4.9648,5.1931)$ for $n=1,2,3,4$ and $5$ respectively. Moreover,  the value of $b_\pi/M$ as $M\to \infty$ is solved as $4.4573$. Consequently, for sufficiently large $M$, we have $b_3'<b_\pi<b_4'$. Therefore, there are at least three discrete bright rings in the BH image for a massive BH. These rings locate at $(b_n+b_n')/(2M)\approx 0.04,\ 1.1$ and $3.5$ respectively with widths $\Delta b/M\approx 0.002,\ 0.3$ and $1.4$.  The above results of analytical calculations are confirmed by numerical computation in their scope of application. For instance, the consistencies of the analytical and numerical results of $b_\pi$ and $b_n'$ with $n=2,3,4$ are shown in Fig. \ref{fig:cribs}. It is also illustrated in Fig. \ref{fig:cribs} that $b_\pi$ becomes smaller than $b_3'$ for small $M$. Hence, there would be two distinguishable bright rings in this case.

%\onecolumngrid

%\twocolumngrid
%\begin{figure}
%\centering
%\includegraphics[width=0.5\textwidth]{shadow2}
%\caption{Overlay }\label{fig:shadow2}
%\end{figure}

\begin{figure}
\centering
\includegraphics[width=0.3\textwidth]{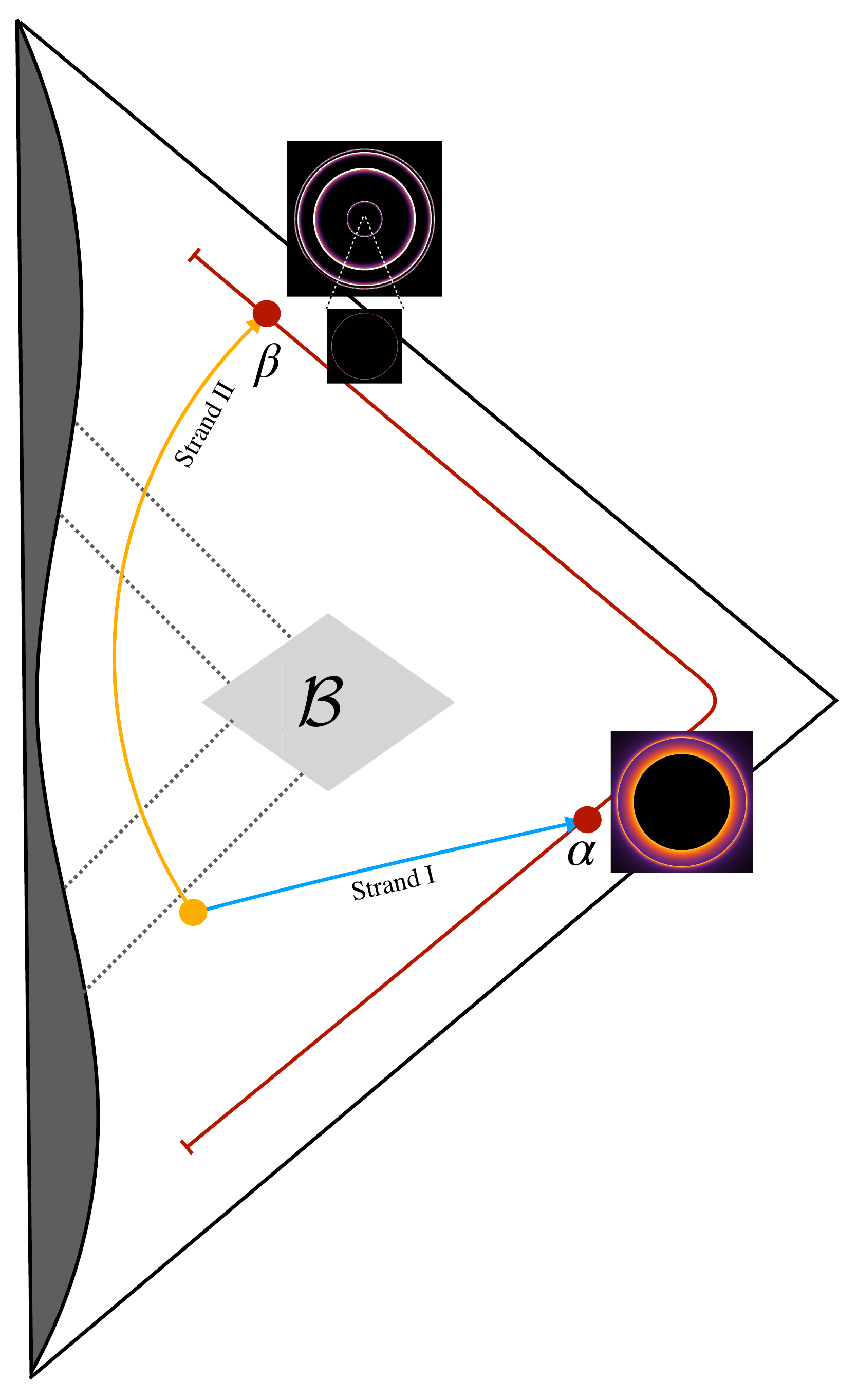}
\caption{The spacetime proposed  in \cite{Han:2023wxg} and its implication for the BH images: The red curve represents the worldine of a static observer situated at $r\gg 2M$. The dashed lines plot the finite inner and outer apparent horizons.  
}\label{fig:movie}
\end{figure}

\section{summary and discussion}
It is a fantastic idea to study possible effects of QG in the images of BHs. However, the QG effects in the studies previous to ours were tiny because they are only concerned with the imaging process of the light rays in the classical region outside the BH horizon. As shown in previous sections, our new idea is to consider the imaging process in the quantum extension of a BH spacetime where the classical singularity can be resolved so that there exists a companion BH $B'$ in the universe earlier than ours, as shown in Fig. \ref{fig:penrose}. This new idea dramatically changed the whole picture, since radiations from the thin disk of $B'$ can enter its horizon, show up from a WH horizon in our universe, and contribute bright rings to the image of the BH $B$, as illustrated in Fig. \ref{fig:shadow}. The positions and widths of these bright rings have been predicted precisely by our calculations. For supermassive BHs, there are at least three discrete bright rings in the BH image, located at $b/(2M) \approx 0.04, 1.1,$ and $3.5$, respectively, with widths $\Delta b/M \approx 0.002, 0.3,$ and $1.4$. These results are universal for a general class of quantum modified spacetimes satisfying Eqs. \eqref{eq:classicallim1} and \eqref{eq:classicallim2}. This discovery opens an experimental window to test the BH-to-WH transition predicted by QG. Moreover, the next-to-leading order term in Eq. \eqref{eq:phiouttot} is model-dependent, which would be encoded in the higher-order correction of the positions and widths of the bright rings and thus provide the information to distinguish different candidate theories.

The idea of observing signals from a companion BH in the universe earlier
than ours, as presented in this paper, can be extended to investigate other models with a similar scenario such as the Bardeen BHs \cite{Ayon-Beato:1998hmi}, the Hayward BHs \cite{Hayward:2005gi} and the renormalization group improved BHs \cite{PhysRevD.62.043008}. It is reasonable to expect that there would also appear some bright rings  in the shadow region of the observed BH. Since numbers, positions, and widths of the rings would be different for different BH models under consideration, the observation can be used to discern these models  and test the different theories. However, all of these BH models, including the one presented in this paper, contain inner Cauchy horizons, similar to the case of a charged Reissner-Nordstr\"om BH.  The existence of a Cauchy horizon may lead to the problem concerning the dynamical stability of the spacetime, particularly due to the mass-inflation effect \cite{PhysRevD.41.1796,PhysRevLett.67.789}.  Nevertheless, even within classical GR, the curvature singularity caused by perturbations is found to be weak and may not necessarily rule out the extension of the spacetime geometry \cite{PhysRevLett.67.789,PhysRevLett.74.1064}. Moreover, quantum effects, such as those arising from the Hawking radiation and QG, are expected to potentially mitigate this instability (see, e.g.,\cite{PhysRevLett.67.789,Bonanno:2022rvo,Carballo-Rubio:2022twq} for various viewpoints on this issue.) 

By utilizing the metric \eqref{eq:metric}, it was recently proposed in \cite{Han:2023wxg} that a new spacetime structure incorporating the QG effects at the end of Hawking evaporation may emerge such that the metric of the new spacetime remains locally the same as Eq. \eqref{eq:metric}, except for a specific region $\mathcal B$. As shown in Fig. \ref{fig:movie},  this $\mathcal B$ region changes the global structure of the spacetime, so that no observer crossing the inner horizon  can receive an infinitely blue-shifted energy from outside. Thus,
this spacetime would not suffer the problem of Cauchy horizon instability. Although the global structure of spacetime was changed by this new proposal, our previous results of BH image would still be valid at the final stage of the observation. To see this, let us analyse the situation illustrated in Fig. \ref{fig:movie}. The light rays emitted from the disk surrounding the outer apparent horizon can split into multiple strands. One one hand, a particular strand  directly reaches the observer at a point $\alpha$ with distant radius $r \gg 2M$. It provides an image resembling the one shown in the bottom right panel of Fig. \ref{fig:shadow}. On the other hand, another strand enters the horizon, traverses the highly quantum region, and eventually reaches the observer at a point $\beta$ in the future. This happens because the metric in that region is locally the same as metric \eqref{eq:metric}. Consequently, the observer receives the image shown in the bottom left panel of Fig. \ref{fig:shadow}  at a latter stage. Therefore, an observer situated at $r \gg 2M$ would witness a movie displaying the entire lifespan of the BH, which begins with the image depicted in the bottom right panel of Fig. \ref{fig:shadow} and concludes with the image shown in the bottom left panel. It is worth noting that for the intermediate episodes, the concrete form of the metric in region $\mathcal B$ is needed in order to study the light strands passing through it.

This work indicates the exciting possibility that the BH image may encode information about QG. It is important to further study this issue in more realistic models, such as the quantum Oppenheimer-Snyder model, where both the emission of matter fields falling into the BH and the absorption of radiation by matter fields should be considered.

%
%It should be noted that the existence of Cauchy horizons in the current model might lead to instability. However, it has been argued that QG effects could change the global causal structure of spacetime and lead to a stable spacetime where the BH-to-WH transition phenomenon would remain \cite{d2021end,Han:2023wxg}. Then, our scheme would still be valid in that case.

As shown in this work, the WH $W$ functions as a projector, taking the image of $B'$ and projecting it onto the image of $B$. This process also allows lights from other sources such as stars in $A'$ to be projected by the WH $W'$. However, just like in our universe, the light from the accretion disk surrounding $B'$ is expected to be much brighter than other sources, and hence the image of $B'$ should be dominant in the final images. Finally, this work opens a new window to detect the effects of QG. It is interesting to note that, in addition to electromagnetic radiations, gravitational waves may also pass through the highly quantum region, bringing on information about QG and potentially being captured by future observers.

\begin{acknowledgments}
C.Z. acknowledges the useful discussion with Hyat Huang, Minyong Guo, and Jinbo Yang at the early stage of this work. This work is supported by the National Natural Science Foundation of China with Grants No. 12275022,  No. 11961131013, and No. 12165005.
\end{acknowledgments}

\onecolumngrid 
\appendix
 \section{Derivation of Eq. \eqref{eq:phiouttot}}\label{app:A}
Given an impact parameter $b=\beta M$, the azimuth angle $\phi(u)$ satisfies
\begin{equation}\label{eq:azimuth}
\begin{aligned}
\frac{\dd\phi}{\dd u}=\frac{1}{\sqrt{-\alpha M^2 u^6+2 M u^3-u^2+\beta^{-2} M^{-2}}}, 
\end{aligned}
\end{equation}
with the initial condition  $\phi(0)=0$. Since $b< b_c$, $G(u)$ has two roots $u_+>0$ and $u_-<0$. They are
\begin{equation}
\begin{aligned}
u_+=\left(\frac{2}{\alpha M}\right)^{1/3}-\frac{1}{6 M}+O\left(M^{-4/3}\right),\ u_-=\frac{x_0}{M}+O(M^{-2})
\end{aligned}
\end{equation} 
where $x_0$ is the real solution to the equation $2 \beta ^2 x^3-\beta ^2 x^2+1=0$. The domain of the Eq. \eqref{eq:azimuth} is $u\in (0,u_+)$. Seeing $M^{-1}$ as a perturbation, one cannot find a single asymptotic series describes $\phi(u)$ uniformly for all values of $u$ in the domain. We  thus need to employ the Matched Asymptotic Expansion (MAE) method. 

For $u\sim u_+$, $u$ is scaled by $x=u( \alpha M/2)^{1/3}$.  Then, Eq. \eqref{eq:azimuth}, up to the leading order, becomes 
\begin{equation}\label{eq:firsteq}
\begin{aligned}
\frac{\dd\phi_{\rm out}}{\dd x}=\left( \frac{\sqrt{\alpha}} {4M}\right)^{1/3} \frac{1}{\sqrt{-x^6+x^3}}.
\end{aligned}
\end{equation}
The solution is 
\begin{equation}
\phi_{\rm out}(x)=c_1-\left(\frac{2\sqrt{\alpha}}{ M}\right)^{1/3}\frac{  _2F_1\left(-\frac{1}{6},\frac{1}{2};\frac{5}{6};x^3\right)}{\sqrt{x}},
\end{equation}
with some integration constant $c_1$ to be fixed.  

We now come to the region $u\sim 0$ where $u$ is scaled by $u=y M^{-1}$. Then, Eq. \eqref{eq:azimuth}, up to the leading order, becomes 
\begin{equation}
\frac{\dd\phi_{\rm in}}{\dd y}=\frac{1}{\sqrt{2  y^3-y^2+\beta^{-2}}}
\end{equation}
The solution to this equation is 
\begin{equation}
\begin{aligned}
\phi_{\rm in}(u)=\frac{\sqrt[4]{3} \sqrt[6]{\Lambda }}{\sqrt[4]{\Lambda ^{2/3}+\Lambda ^{4/3}+1}}\left[F\left(\arccos(\frac{a_+}{a_-}),k\right)-F\left(\arccos(\frac{y-a_+}{y-a_-}),k\right)\right]
\end{aligned}
\end{equation}
where the initial condition $\phi(0)=0$ is applied, $F(\alpha,k)$ denotes the first kind elliptic integral
\begin{equation}
F(\alpha,k)=\int_0^\alpha\frac{\dd\theta}{\sqrt{1-k^2\sin^2\theta}},
\end{equation}
and
\begin{equation}
\begin{aligned}
\Lambda=&-\frac{6 \left(\sqrt{81-3 \beta ^2}-9\right)}{\beta ^2}-1,\quad  k=\frac{1}{2} \left(\frac{\sqrt{3} \left(\Lambda ^{2/3}+1\right)}{\sqrt{\Lambda ^{2/3}+\Lambda ^{4/3}+1}}+2\right)^{1/2},\\
a_{\pm}=&\frac{-\Lambda ^{2/3}-1+\Lambda^{1/3}\pm \sqrt{3} \sqrt{\Lambda^{4/3}+\Lambda^{2/3}+1 }}{6\sqrt[3]{\Lambda }}.
\end{aligned}
\end{equation}

By Eq. \eqref{eq:azimuth}, the inner solution $\phi_{\rm in}$ is valid for $\alpha M^2 u^6\ll  M u^3$, i.e., $u\ll (\alpha M)^{-1/3}$, and the outer solution is valid for $M u^3\gg u^2$, i.e., $u\gg M^{-1}$. We thus have the intermedia region $M^{-1}\ll u\ll  (\alpha M)^{-1/3}$, where the two solutions $\phi_{\rm in}$ and $\phi_{\rm out}$ match. We thus introduce $u=c M^{-p}$ with $1/3<p<1$ so that $c M^{-p}$ lies in the intermedia region. The matching leads to
\begin{equation}
\lim_{M\to\infty}\phi_{\rm in}\left(c \alpha^{1/3} M^{1/3-p}/2^{1/3} \right)=\lim_{M\to\infty}\phi_{\rm out}(c M^{1-p}). 
\end{equation}
which fixes $c_1$ to be
\begin{equation}
c_1=\frac{\sqrt[4]{3} \sqrt[6]{\Lambda }}{\sqrt[4]{\Lambda ^{2/3}+\Lambda ^{4/3}+1}}F\left( \arccos(\frac{a_+}{a_-}),k\right)=\int_0^\infty\frac{\dd y}{\sqrt{2y^3-y^2+\beta^{-2}}}
\end{equation}
A uniformly valid approximation throughout the entire domain is obtained by adding the two expansion and subtracting off the common behavior. Therefore, we have
\begin{equation}\label{eq:resultphi}
\begin{aligned}
\phi(u)
=&\frac{\sqrt{2}}{\sqrt{Mu}} - \frac{\sqrt{2}\ _2F_1\left(-\frac{1}{6},\frac{1}{2};\frac{5}{6};\frac{\alpha M}{2}u^3\right)}{\sqrt{Mu}}- \frac{\sqrt[4]{3} \sqrt[6]{\Lambda }}{\sqrt[4]{\Lambda ^{2/3}+\Lambda ^{4/3}+1}}F\left(\arccos(\frac{Mu-a_+}{Mu-a_-},k)\right)\\
&+\int_0^\infty\frac{\dd y}{\sqrt{2 y^3-y^2+\beta^{-2}}}
\end{aligned}
\end{equation} 

For the total change of the azimuth angle $\phi_{\rm tot}(\beta M)=2\phi(u_+)$, a straight forward calculation gives
\begin{equation}
\begin{aligned}
\frac{\phi_{\rm tot}(\beta M)}{2}=\int_0^\infty\frac{\dd y}{\sqrt{2 y^3-y^2+\beta^{-2}}}-\frac{  \sqrt[3]{2} \sqrt[6]{\alpha } \sqrt{\pi }\,  \Gamma \left(\frac{5}{6}\right)}{\Gamma \left(\frac{1}{3}\right)}\sqrt[3]{\frac{1}{M}}+O((\sqrt\alpha/M)^{2/3}).
\end{aligned}
\end{equation}
For $\phi_{\rm out}(\beta M)=\phi(1/r_+)$, we have
\begin{equation}
\begin{aligned}
\phi_{\rm out}(\beta M)=\int_{0}^{1/2}\frac{\dd y}{\sqrt{2y^3-y^2+\beta^{-2}}}+O(\sqrt\alpha/M),
\end{aligned}
\end{equation}
where $r_+=2M(1+O(M^{-1}))$ is applied.

 \section{Derivation of the analytical approximation in Fig. \ref{fig:phivalues}}\label{app:B}
 We are concerned with the differential equation 
 \begin{equation}\label{eq:phiu}
 \begin{aligned}
 \frac{\dd \phi}{\dd u}=\frac{1}{\sqrt{G(u,b_c-\epsilon^2)}},
 \end{aligned}
 \end{equation}
 for $\phi(u)$, with the initial condition $\phi(0)=0$ and $\epsilon\ll \sqrt{b_c}$.  The "boundary layer" exists at $w_c$ with $w_c$ being the double root of $G(w,b_c)$. Since $\phi_{\rm tot}(b)=2\phi(u_+)$ with $u_+>u_c$, we need to divide the entire calculation into two part. The first part considers the solution $\phi(u)$ for $u\leq u_c$ to get $\phi(u_c)$. The second part concerns the solution $\phi(u)$  for $u\geq u_c$ where $\phi(u_c)$ obtained from the previous calculation performs as the initial condition. 
 
 \subsection{solution for $u<u_c$}
 Let us introduce $v=u_c-u$. Then, one gets
 \begin{equation}
 G(u_c-v,b)=\sum_{n=2}^6 \frac{1}{n!}\partial_u^{n}G(u_c,b_c)v^n+\frac{1}{(b_c-\epsilon)^2}-\frac{1}{b_c^2}. 
 \end{equation}
 For $v=u_c-u\gg \sqrt{|\partial_u^2 G(u_c,b_c)|/b_c^3}\, \epsilon$, the equation Eq. \eqref{eq:phiu} can be expand directly at $\epsilon=0$. We get 
 \begin{equation}\label{eq:diff}
 \frac{\dd\phi_{\rm out}}{\dd u}=\frac{1}{\sqrt{G(u,b_c)}}. 
 \end{equation}
 Let $v_1$, $v_2$ and $v_3\pm i v_4$ be the remaining four roots besides $0$ of $G(u_c-v,b_c)$ with 
 \begin{equation}
 v_2<0<u_c<v_1.
 \end{equation}
 Clearly, the turning point $u_+$ relates to $v_2$ by $u_+=u_c-v_2+O(\epsilon)$.   We introduce the reals numbers $A_1$, $A_2$, $B_1$, $B_2$, $\lambda$ and $\beta$ with $A_1<0$ such that
\begin{equation}\label{eq:ABAB}
(v_1-v)(v-v_2)=A_1(v-\lambda)^2+B_1(v-\beta)^2,\quad (v-v_3-i v_4)(v-v_3+iv_4)=A_2(v-\lambda)^2+B_2(v-\beta)^2.
\end{equation}
Indeed, the condition $A_1<0$ leads to $B_1,A_2,B_2>0$. Then, a straightforwards calculation shows that
\begin{equation}
\begin{aligned}
\phi_{\rm out}(u)=&\frac{k}{\sqrt{\alpha}M (\lambda-\beta)\beta\sqrt{A_2B_1}}\int^{\arccos(\sqrt{\frac{-A_1}{B_1}}\frac{u_c-u-\lambda}{u_c-u-\beta})}_0 \frac{\dd \theta}{\sqrt{1-k^2\sin^2\theta}} \\
&-\frac{k n\lambda A_1}{ \sqrt{\alpha}M \beta^3 B_1\sqrt{A_2B_1}} \int^{\arccos(\sqrt{\frac{-A_1}{B_1}}\frac{u_c-u-\lambda}{u_c-u-\beta})}_0 \frac{1 }{1 -n\sin^2(\theta)}\frac{\dd \theta}{\sqrt{1-k^2\sin^2\theta}}\\
&+\frac{1}{2\sqrt{\alpha}M}L\left(\frac{u_c-u-\lambda}{u_c-u-\beta}\right)+C_f^+,
\end{aligned}
\end{equation}
where $C_f^+$ is the integration constant, the first two terms involve the elliptic integrals of the first and the third kinds respectively with
\begin{equation}
k=\sqrt{\frac{A_2B_1}{A_2B_1-A_1B_2}}<1,\quad n=\frac{\beta^2B_1}{\lambda^2 A_1+\beta^2B_1}>1.
\end{equation}
and  the function $L$ reads
\begin{equation}
\begin{aligned}
L(t)=\frac{\ln \left(\left|\frac{\sqrt{A_2 t^2+B_2} \sqrt{A_1 \lambda ^2+\beta ^2 B_1}+\sqrt{A_1 t^2+B_1} \sqrt{A_2 \lambda ^2+\beta ^2 B_2}}{\sqrt{A_2 t^2+B_2} \sqrt{A_1 \lambda ^2+\beta ^2 B_1}-\sqrt{A_1 t^2+B_1} \sqrt{A_2 \lambda ^2+\beta ^2 B_2}}\right|\right)}{\sqrt{A_1 \lambda ^2+\beta ^2 B_1} \sqrt{A_2 \lambda ^2+\beta ^2 B_2}}.
\end{aligned}
\end{equation}
In the above derivations, we applied $B_1\beta^2+A_1\lambda^2=-v_1v_2>0$. Moreover,  it could happen that, in the second integral  of $\phi_{\rm out}(u)$, $\theta_0$ satisfying $1-n \sin^2\theta$ is contained in the integration interval. In this case, the Cauchy principal value is chosen, as usually done in defining the elliptic integral. Indeed, the singularity of this integral occurs at $u=u_c$. Hence, if for a $u_o<u_c$ the singularity $\theta_0$  is contained in the integration interval $(0,\arccos(\sqrt{\frac{-A_1}{B_1}}\frac{u_c-u-\lambda}{u_c-u-\beta}))$, the singularity will appear to lie in $(0,\arccos(\sqrt{\frac{-A_1}{B_1}}\frac{u_c-u-\lambda}{u_c-u-\beta}))$ for all $u<u_c$.

Now let us come to the inner solution  $\phi_{\rm in}(u)$ for $u\sim u_c$. For $u_c-u \ll |\partial_u^2 G(u_c,b_c)/\partial_u^3 G(u_c,b_c)|$, we scale the variable $v=u_c-u$ by 
\begin{equation}
u_c-u=y\epsilon.
\end{equation}
Then, expanding Eq. \eqref{eq:diff} up to the leading order, one gets
\begin{equation}
\frac{\dd \phi_{\rm in}}{\dd y}=-\frac{1}{\sqrt{\alpha}M\sqrt{y^2 \left(A_1 \lambda ^2+\beta ^2 B_1\right) \left(A_2 \lambda ^2+\beta ^2 B_2\right)+\frac{2}{b_c^3}}}
\end{equation}
whose solution is 
\begin{equation}
\begin{aligned}
\phi_{\rm in}(y)=- \frac{\mathrm{arcsinh}\left(\sqrt{b_c^3\left(A_1 \lambda ^2+\beta ^2 B_1\right) \left(A_2 \lambda ^2+\beta ^2 B_2\right) /2 }\, y\right)}{\sqrt{\alpha}M\sqrt{\left(A_1 \lambda ^2+\beta ^2 B_1\right) \left(A_2 \lambda ^2+\beta ^2 B_2\right) }}+C_I.
\end{aligned}
\end{equation}

The two solutions $\phi_{\rm in}$ and $\phi_{\rm out}$ match in the intermedia region, which leads to 
\begin{equation}\label{eq:mathchingccondition}
\begin{aligned}
\lim_{\epsilon\to 0}\phi_{\rm in}(c\epsilon^{p-1} )=\lim_{\epsilon\to 0} \phi_{\rm out}(u_c-c\epsilon^p),
\end{aligned}
\end{equation}
with $0<p<1$. For the left hand side of Eq. \eqref{eq:mathchingccondition}, we have
\begin{equation}
\begin{aligned}
\phi_{\rm in}(c\epsilon^{p-1})=-\frac{\ln \left(2\sqrt{\alpha M^2b_c^3\left(A_1 \lambda ^2+\beta ^2 B_1\right) \left(A_2 \lambda ^2+\beta ^2 B_2\right) /2 }\right)+\ln(c\epsilon^{p-1})}{\sqrt{\alpha}M\sqrt{\left(A_1 \lambda ^2+\beta ^2 B_1\right) \left(A_2 \lambda ^2+\beta ^2 B_2\right) }}+C_I. 
\end{aligned}
\end{equation}
The right hand side of Eq. \eqref{eq:mathchingccondition} is more complicated to be dealt with. We need to employ the following formula for the elliptic integrals $\Pi(n;\phi, k)$ (see Eq. (19.7.8) in \cite{NIST:DLMF} ), 
\begin{equation}
\begin{aligned}
\Pi(n;\phi, k)+\Pi(\omega;\phi, k)=F(\phi, k)+\sqrt{\xi}R_C((\xi-1)(\xi-k^2),(\xi-n)(\xi-\omega))
\end{aligned}
\end{equation}
where $\phi\in [0,\pi/2]$, $n \omega=k^2$, $\xi=1/\sin^2(\phi)$ and 
\begin{equation}
R_C(x,y)=\left\{
\begin{aligned}
\frac{1}{\sqrt{y-x}}\arccos(\sqrt{x/y}),\ 0\leq x<y,\\
\frac{1}{\sqrt{x-y}}\ln(\frac{\sqrt{x}+\sqrt{x-y}}{\sqrt{y}}),\ 0<y< x,\\
\frac{1}{\sqrt{x-y}}\ln(\frac{\sqrt{x}+\sqrt{x-y}}{\sqrt{-y}}),\ y<0< x.
\end{aligned}
\right.
\end{equation}
By the definition of $R_C$, we get for $|y|\ll 1$
\begin{equation}
R_C(x,y)=\frac{1}{2\sqrt{x}}\left(\ln(4x)-\ln(y)\right)+O(y).
\end{equation}
Note that our convention defines $\Pi(n;\phi,k)=2\Pi(n;\pi/2,k)-\Pi(n;\pi-\phi,k)$ for $\pi\in(\pi/2,\pi)$ so that $\Pi(n;\phi,k)$ is $C^1$ at $\phi=\pi/2$.

Taking advantage of these formulas, a straightforward calculation shows 
\begin{equation}
\begin{aligned}
\phi_{\rm out}(u_c-c\epsilon^p)
=&\frac{k}{\sqrt\alpha M (\lambda-\beta)\beta\sqrt{A_2B_1}} F\left(\arccos(\sqrt{\frac{-A_1}{B_1}}\frac{\lambda}{\beta}), k\right)\\
&-\frac{kn\lambda A_1}{ \sqrt\alpha M \beta^3 \sqrt{A_2B_1}}\Bigg\{2\Theta(-\lambda\beta)\Pi(n,\pi/2|k)-\sgn(\lambda\beta)\Pi\left(\frac{k^2}{n},\arccos(\sqrt{\frac{-A_1}{B_1} } \big|\frac{\lambda}{\beta}\big|),k \right)\\
&+\sgn(\lambda\beta)F\left(\arccos(\sqrt{\frac{-A_1}{B_1} } \big|\frac{\lambda}{\beta}\big|),k \right)\Bigg\} +\frac{
\ln(\frac{4 \left(A_1 \lambda^2+B_1 \beta^2\right)^2 \left(A_2 \lambda^2+B_2 \beta^2\right)^2}{(\beta-\lambda)^2 (A_2 B_1-A_1 B_2) \left|\left(A_1 A_2 \lambda^4+2 A_2 B_1 \beta^2 \lambda^2+B_1 B_2 \beta^4\right)\right|})
}{2\sqrt{\alpha}M \sqrt{A_1 \lambda ^2+\beta ^2 B_1} \sqrt{A_2 \lambda ^2+\beta ^2 B_2} }\\
&-\frac{\ln(c\epsilon^p)}{\sqrt{\alpha}M\sqrt{  \left(A_2 \lambda ^2+\beta ^2 B_2\right)\left(A_1 \lambda ^2+\beta ^2 B_1\right)}}+C_f^+,
\end{aligned}
\end{equation}
with $\Theta$ denoting the step function.  Thus, the matching Eq. \eqref{eq:mathchingccondition}
\begin{equation}\label{eq:CI}
\begin{aligned}
 C_I=&C_f^++\frac{k}{\sqrt{\alpha}M(\lambda-\beta)\beta\sqrt{A_2B_1}}F\left(\arccos(\sqrt{\frac{-A_1}{B_1}}\frac{\lambda}{\beta}), k\right)\\
&-\frac{kn\lambda A_1}{ \sqrt{\alpha}M\beta^3\sqrt{A_2B_1} }\Bigg\{2\Theta(-\lambda\beta)\Pi(n,\pi/2,k)-\sgn(\lambda\beta)\Pi\left(\frac{k^2}{n},\arccos(\sqrt{\frac{-A_1}{B_1} } \big|\frac{\lambda}{\beta}\big|),k \right)\\
&+\sgn(\lambda\beta)F\left(\arccos(\sqrt{\frac{-A_1}{B_1} } \big|\frac{\lambda}{\beta}\big|),k \right)\Bigg\}+\frac{
\ln(\left|\frac{8\alpha M b_c^3 \left(A_1 \lambda^2+B_1 \beta^2\right)^3 \left(A_2 \lambda^2+B_2 \beta^2\right)^3}{(\beta-\lambda)^2 (A_2 B_1-A_1 B_2) \left(A_1 A_2 \lambda^4+2 A_2 B_1 \beta^2 \lambda^2+B_1 B_2 \beta^4\right)}\right|)
}{2\sqrt{\alpha}M \sqrt{A_1 \lambda ^2+\beta ^2 B_1} \sqrt{A_2 \lambda ^2+\beta ^2 B_2} }\\
&+\frac{\ln(\epsilon^{-2}M)}{2\sqrt{\alpha}M\sqrt{\left(A_1 \lambda ^2+\beta ^2 B_1\right) \left(A_2 \lambda ^2+\beta ^2 B_2\right) }},
\end{aligned}
\end{equation}
where it should be noted that
\begin{equation}
\sqrt{\alpha}M\sqrt{\left(A_1 \lambda ^2+\beta ^2 B_1\right) \left(A_2 \lambda ^2+\beta ^2 B_2\right)} =\sqrt{\partial_u^2G(u_c,b_c)/2}.
\end{equation}
For $C_f^+$, the initial condition $\phi_{\rm out}(0)=0$ leads to
\begin{equation}
\begin{aligned}
C_f^+=&-\frac{k}{\sqrt{\alpha}M (\lambda-\beta)\beta\sqrt{A_2B_1}}F\left(\arccos(\sqrt{\frac{-A_1}{B_1}}
\frac{u_c-\lambda}{u_c-\beta}),k \right) \\
&+\frac{k n\lambda A_1}{ \sqrt{\alpha}M \beta^3 B_1\sqrt{A_2B_1}} \Pi\left(n;\arccos(\sqrt{\frac{-A_1}{B_1}}\frac{u_c-\lambda}{u_c-\beta}),k\right)-\frac{1}{2\sqrt{\alpha}M}L\left(\frac{u_c-\lambda}{u_c-\beta}\right).
\end{aligned}
\end{equation}
%Finally, the solution $\phi(u)$ for $u\leq u_c$ reads
%\begin{equation}
%\begin{aligned}
%\phi(u)=&\frac{k}{\sqrt{\alpha}M(\lambda-\beta)\beta\sqrt{A_2B_1}}\int^{\arccos(\sqrt{\frac{-A_1}{B_1}}\frac{u_c-u-\lambda}{u_c-u-\beta})}_0 \frac{\dd \theta}{\sqrt{1-k^2\sin^2\theta}} \\
%&-\frac{kn\lambda A_1}{ \sqrt{\alpha}M\beta^3\sqrt{A_2B_1} }\int^{\arccos(\sqrt{\frac{-A_1}{B_1}}\frac{u_c-u-\lambda}{u_c-u-\beta})}_0 \frac{1 }{1 -n\sin^2(\theta)}\frac{\dd \theta}{\sqrt{1-k^2\sin^2\theta}}+\frac{1}{2\sqrt{\alpha}M}L(u_c-u)\\
%&- \frac{\mathrm{arcsinh}\left(\sqrt{b_c^3\left(A_1 \lambda ^2+\beta ^2 B_1\right) \left(A_2 \lambda ^2+\beta ^2 B_2\right) /2 }\, (u_c-u)/\epsilon\right)}{\sqrt{\alpha}M\sqrt{\left(A_1 \lambda ^2+\beta ^2 B_1\right) \left(A_2 \lambda ^2+\beta ^2 B_2\right) }}\\
%&+\frac{\ln \left(2\sqrt{b_c^3\left(A_1 \lambda ^2+\beta ^2 B_1\right) \left(A_2 \lambda ^2+\beta ^2 B_2\right) /2 }\right)+\ln((u_c-u)/\epsilon)}{\sqrt{\alpha}M\sqrt{\left(A_1 \lambda ^2+\beta ^2 B_1\right) \left(A_2 \lambda ^2+\beta ^2 B_2\right) }}+C_f^+.
%\end{aligned}
%\end{equation}
For $u=u_c$, one has
\begin{equation}
\begin{aligned}
\phi(u_c)=\phi_{\rm in}(0)=C_I. 
\end{aligned}
\end{equation}
\subsection{solution for $u>u_c$}
Now, let us consider $u>u_c$. In this region, the inner solution is 
\begin{equation}
\begin{aligned}
\phi_{\rm in}(u)=- \frac{\mathrm{arcsinh}\left(\sqrt{\alpha M^2b_c^3\left(A_1 \lambda ^2+\beta ^2 B_1\right) \left(A_2 \lambda ^2+\beta ^2 B_2\right) /2 }\, (u_c-u)\epsilon^{-1}\right)}{\sqrt{\alpha}M\sqrt{\left(A_1 \lambda ^2+\beta ^2 B_1\right) \left(A_2 \lambda ^2+\beta ^2 B_2\right) }}+C_I.
\end{aligned}
\end{equation}
where $C_I$ is given by Eq. \eqref{eq:CI}. For the outer solution, we have
\begin{equation}
\begin{aligned}
\phi_{\rm out}(u)=&-\frac{k}{\sqrt{\alpha}M (\lambda-\beta)\beta\sqrt{A_2B_1}}\int^{\arccos(\sqrt{\frac{-A_1}{B_1}}\frac{u_c-u-\lambda}{u_c-u-\beta})}_0 \frac{\dd \theta}{\sqrt{1-k^2\sin^2\theta}} \\
&+\frac{k n\lambda A_1}{ \sqrt{\alpha}M \beta^3 B_1\sqrt{A_2B_1}} \int^{\arccos(\sqrt{\frac{-A_1}{B_1}}\frac{u_c-u-\lambda}{u_c-u-\beta})}_0 \frac{1 }{1 -n\sin^2(\theta)}\frac{\dd \theta}{\sqrt{1-k^2\sin^2\theta}}\\
&-\frac{1}{2\sqrt{\alpha}M}L\left(\frac{u_c-u-\lambda}{u_c-u-\beta}\right)+C_f^-. 
\end{aligned}
\end{equation}
Repeating the calculation for $u<u_c$, one get by matching the two solutions in the intermedia region, 
\begin{equation}
\begin{aligned}
C_f^-=&2C_I-C_f^+.
\end{aligned}
\end{equation}
Then, the entire solution for $u>u_c$ is 
\begin{equation}
\begin{aligned}
\phi(u)=&\frac{-k}{ \sqrt{\alpha}M (\lambda-\beta)\beta\sqrt{A_2B_1}}\int^{\arccos(\sqrt{\frac{-A_1}{B_1}}\frac{u_c-u-\lambda}{u_c-u-\beta})}_0 \frac{\dd \theta}{\sqrt{1-k^2\sin^2\theta}} \\
&+\frac{kn\lambda A_1}{ \sqrt{\alpha}M \beta^3 B_1\sqrt{A_2B_1}}\int^{\arccos(\sqrt{\frac{-A_1}{B_1}}\frac{u_c-u-\lambda}{u_c-u-\beta})}_0 \frac{1 }{1 -n\sin^2(\theta)}\frac{\dd \theta}{\sqrt{1-k^2\sin^2\theta}}\\
&-\frac{1}{2\sqrt{\alpha}M}L\left(\frac{u_c-u-\lambda}{u_c-u-\beta}\right)- \frac{\mathrm{arcsinh}\left(\sqrt{\alpha M^2 b_c^3\left(A_1 \lambda ^2+\beta ^2 B_1\right) \left(A_2 \lambda ^2+\beta ^2 B_2\right) /2 }\, (u_c-u)/\epsilon\right)}{\sqrt{\alpha}M\sqrt{\left(A_1 \lambda ^2+\beta ^2 B_1\right) \left(A_2 \lambda ^2+\beta ^2 B_2\right) }}\\
&+\frac{\ln \left(2\sqrt{\alpha M^2 b_c^3\left(A_1 \lambda ^2+\beta ^2 B_1\right) \left(A_2 \lambda ^2+\beta ^2 B_2\right) /2 }\right)+\ln((u_c-u)/\epsilon)}{\sqrt{\alpha}M\sqrt{\left(A_1 \lambda ^2+\beta ^2 B_1\right) \left(A_2 \lambda ^2+\beta ^2 B_2\right) }}-C_f^++2C_I.
\end{aligned}
\end{equation}
The total azimuth angle is $2\phi(u_+)=2\phi(u_c-v_2+O(\epsilon))$. We thus have
\begin{equation}
\begin{aligned}
\frac{\phi_{\rm tot}(b_c-\epsilon^2)}{2}=&-C_f^++2C_I.
\end{aligned}
\end{equation}
where we used
\begin{equation}
\sqrt{\frac{-A_1}{B_1}}\frac{v_2-\lambda}{v_2-\beta})=1. 
\end{equation}
For $\phi_{\rm out}(b_c-\epsilon^2)$, let $v_+=u_c-1/r_+$ with $r_+$ denoting the radius of the event horizon. We have
\begin{equation}
\begin{aligned}
\phi_{\rm out}(b_c-\epsilon^2)=&2\phi_{\rm tot}-\phi(u_c-v_+)\\
=&2C_I-C_f^++\frac{k}{ \sqrt{\alpha}M (\lambda-\beta)\beta\sqrt{A_2B_1}}F\left(\arccos(\sqrt{\frac{-A_1}{B_1}}\frac{v_+-\lambda}{v_+-\beta}),k\right) \\
&-\frac{kn\lambda A_1}{ \sqrt{\alpha}M \beta^3 B_1\sqrt{A_2B_1}}\Pi\left(n; \arccos(\sqrt{\frac{-A_1}{B_1}}\frac{v_+-\lambda}{v_+-\beta}),k\right)+\frac{1}{2\sqrt{\alpha}M}L\left(\frac{v2-\lambda}{v2-\beta}\right).
\end{aligned}
\end{equation}

\bibliography{reference}

\end{document}